\documentclass[12pt]{article}
\usepackage{psfig}
\begin{document}

\def\lesssim{\mathrel{\mathpalette\vereq<}}
\def\gtrsim{\mathrel{\mathpalette\vereq>}}
\makeatletter
\def\vereq#1#2{\lower3pt\vbox{\baselineskip1.5pt \lineskip1.5pt
\ialign{$\m@th#1\hfill##\hfil$\crcr#2\crcr\sim\crcr}}}
\makeatother

\newcommand{\rem}[1]{{\bf #1}}
\newcommand{\gev}{{\rm GeV}}
\newcommand{\mev}{{\rm MeV}}
\newcommand{\kev}{{\rm keV}}
\newcommand{\ev}{{\rm eV}}
\newcommand{\cm}{{\rm cm}}
\newcommand{\mpl}{M_{Pl}}
\def\pl#1#2#3{{\it Phys. Lett. }{\bf B#1~}(#2)~#3}
\def\zp#1#2#3{{\it Z. Phys. }{\bf C#1~}(#2)~#3}
\def\prl#1#2#3{{\it Phys. Rev. Lett. }{\bf #1~}(#2)~#3}
\def\rmp#1#2#3{{\it Rev. Mod. Phys. }{\bf #1~}(#2)~#3}
\def\prep#1#2#3{{\it Phys. Rep. }{\bf #1~}(#2)~#3}
\def\pr#1#2#3{{\it Phys. Rev. }{\bf D#1~}(#2)~#3}
\def\np#1#2#3{{\it Nucl. Phys. }{\bf B#1~}(#2)~#3}
\def\d#1{\left[ #1 \right]_D}
\def\f#1{\left[ #1 \right]_F}
\def\a#1{\left[ #1 \right]_A}
\def\VEV#1{\left\langle #1\right\rangle}
\let\vev\VEV

\renewcommand{\thefootnote}{\fnsymbol{footnote}}
\setcounter{footnote}{0}
\begin{titlepage}
\begin{center}
\hfill    UCB-PTH-00/18\\                                                      %
\hfill    LBNL-46180\\                                                         %
\hfill    hep-ph/0006312\\
\hfill    \today \\%
\vskip .5in
{\Large \bf Small Neutrino Masses\\from Supersymmetry Breaking
}
\vskip .50in
Nima Arkani-Hamed, Lawrence Hall, Hitoshi Murayama, David Smith and
Neal Weiner
\vskip 0.05in
{\em Department of Physics\\
University of California, Berkeley, California 94720}                     %
\vskip 0.05in
and
\vskip 0.05in
{\em Theoretical Physics Group\\
Ernest Orlando Lawrence Berkeley National Laboratory\\                    %
University of California, Berkeley, California 94720}                     %
\vskip .5in
\end{center}
\vskip .5in
\begin{abstract}
An alternative to the conventional see-saw mechanism is proposed to
explain the origin of small neutrino masses in supersymmetric
theories. The masses and couplings of the right-handed neutrino field
are suppressed by supersymmetry breaking, in a way similar to the
suppression of the Higgs doublet mass, $\mu$. New mechanisms for light
Majorana, Dirac and sterile neutrinos arise, depending on the degree
of suppression. Superpartner phenomenology is greatly altered by the
presence of weak scale right-handed sneutrinos, which may have a
coupling to a Higgs boson and a left-handed sneutrino. The sneutrino
spectrum and couplings are quite unlike the conventional case - the
lightest sneutrino can be the dark matter and predictions are given
for event rates at upcoming halo dark matter direct detection
experiments. Higgs decays and search strategies are changed. Copious
Higgs production at hadron colliders can result from cascade decays of
squarks and gluinos.
\end{abstract}                                                                 %
\end{titlepage}                                                                %
\renewcommand{\thepage}{\arabic{page}}                                         %
\setcounter{page}{1}                                                           %
\renewcommand{\thefootnote}{\arabic{footnote}}                                 %
\setcounter{footnote}{0}                                                       %
\setcounter{footnote}{0}
\section{Introduction}
Why are the neutrinos much lighter than the charged leptons, but not 
absolutely massless? It is universally recognized that this can be simply 
and elegantly understood in an $SU(2) \times U(1)$ effective 
theory. The most general, gauge-invariant interactions of dimension less 
than six, which can lead to masses for the known leptons from the 
vacuum expectation value (vev) of a Higgs doublet, are:
\begin{equation}
\mathcal{L}_{eff} = \lambda LEH + {\lambda' \over M} LLHH
\label{eq:leff}
\end{equation}
where $L$ and $E$ are the lepton doublet and singlet fields, and $H$ is the 
Higgs doublet. The dimensionless matrix of Yukawa couplings, $\lambda$, 
has a hierarchy of eigenvalues to describe the masses of the charged 
leptons. Such a hierarchy could result by promoting the couplings to 
fields, $\lambda \rightarrow \phi/M$, which acquire vevs to sequentially, 
spontaneously break the flavor symmetry $G_F$. Such a flavor 
symmetry will also result in a certain structure for the neutrino mass 
matrix via $\lambda'$. The mass scale $M$ is the cutoff of the low-energy 
effective theory. The crucial point is that if this cutoff is very 
large, for example the Planck or gauge coupling unification scale, then 
the neutrino masses are very small. While the charged lepton masses are 
linear in the Higgs vev, $v$, the neutrino masses are quadratic:
\begin{equation}
m_\nu \approx {v^2 \over M}.
\label{eq:mnu}
\end{equation}
The power of this effective field theory approach is that no assumption 
need be made about the full theory at or above the scale $M$. The only 
assumption is that the low-energy theory is the most general allowed by 
its symmetries. Nevertheless, there is a very simple 
theory which does lead to the dimension 5 operator of (\ref{eq:leff}). 
Right-handed neutrino fields, $N$, are introduced, with Majorana masses $M$ and 
couplings to the lepton doublets:
\begin{equation}
\mathcal{L} = {M \over 2} NN + \xi LNH
\label{eq:l}
\end{equation}
where $M$ and $\xi$ are mass matrices. 
On integrating out the heavy neutrinos, the well-known see-saw mechanism 
gives $\lambda'/M$ in Eq.~(\ref{eq:leff}) by $\xi^T M^{-1} \xi$ \cite{seesaw}.

In supersymmetric theories, there is a very important reason for questioning 
this simple view of neutrino masses: the low-energy effective theory must 
contain more fields than the leptons, Higgs boson and their superpartners. In 
particular, there are two Higgs doublet superfields, $H_u$ and $H_d$, and 
there is another sector of the theory which spontaneously breaks 
supersymmetry, and triggers electroweak symmetry breaking. 
At first sight these additions seem irrelevant to the question of 
neutrino masses, but closer inspection reveals new opportunities. An 
important objection to the minimal supersymmetric standard model is that 
it is not the most general low-energy effective theory consistent with 
$SU(3) \times SU(2) \times U(1)$ gauge symmetry. The gauge symmetry 
allows a bare mass term for the Higgs fields, $[\mu H_u H_d]_F$, giving 
the expectation  $\mu \approx M$, which removes the Higgs from the 
low-energy theory. This is known as the ``$\mu$ problem'' in
supersymmetric theories.

In this paper we take supersymmetry to be broken in a hidden sector,
at the intermediate scale $m_I$, via 
fields $Z$: $\vev{F_Z} \approx m_I^2 \approx v M_{Pl}$, where $M_{Pl}$ 
is the Planck mass. The supersymmetry breaking is communicated to the 
standard model by supergravitational interactions, so that the cutoff for 
the low-energy effective field theory is $M_{Pl}$.
This gives rise to the superpartner masses at the weak scale in the
usual way.
In this case, the above ``$\mu$ problem'' is easily solved by introducing a 
further global symmetry, $G$, and dividing the light matter superfields into 
two types. There are fields which are chiral with respect to the gauge interactions, 
such as $L$ and $E$, which 
are guaranteed to be massless until the gauge symmetry is broken, and 
there are fields which are vector-like, such as $H_u + H_d$, which are 
kept massless only via $G$. Furthermore, the fields $\phi$ which break 
the flavor symmetry do not break $G$ --- the vector-like 
fields acquire mass only from supersymmetry breaking. 
There is a subset of the $Z$ fields, which we call $X$,
that transform non-trivially under $G$.
This ensures that $\mu$ is of order the 
supersymmetry breaking scale, as it must be for electroweak symmetry 
breaking to occur successfully \cite{hall,giudicemasiero}. 
In particular, the operator
\begin{equation}
{ 1 \over M_{Pl}} [X^\dagger H_u H_d]_D
\label{eq:mu}
\end{equation}
leads to $\mu \approx F / M_{Pl} \approx m_I^2/M_{Pl} \approx v$,
where $F$ is the vev of the highest component of $X$.

The right-handed neutrino fields, $N$, like $H_u + H_d$, are vector-like with respect to the gauge 
interactions, hence the crucial question becomes how they transform under 
$G$. If they are also vector-like with respect to $G$, they will acquire a large mass, whence the 
see-saw proceeds via (\ref{eq:l}) as usual. Alternatively, they may be 
protected from acquiring a large mass by $G$, in which case they will 
appear in the low-energy effective theory. The question of neutrino 
masses now becomes much richer, since it is necessary to study all 
possible interactions of $L$ and $N$ in the low-energy theory. In 
particular, mass terms can be induced via interactions with Higgs vevs and 
with $X$ vevs 

By analogy with (\ref{eq:mu}), the right-handed neutrinos may acquire 
mass via the operator
\begin{equation}
{ 1 \over M_{Pl}} [X^\dagger NN]_D
\label{eq:nn}
\end{equation}
giving the right-handed neutrinos a Majorana mass of order the weak scale, $v$. 
The see-saw mechanism can still be operative if the Yukawa couplings are 
suppressed. For example, if one of the $X$ fields acquires an $A$
component vev at the intermediate scale, this occurs via
\begin{equation}
{ 1 \over M_{Pl}} [XLNH_u]_F.
\label{eq:xlnh}
\end{equation}
In general, the coefficients of these higher dimensional operators are 
understood to depend on flavor symmetry breaking, and are functions of 
$\phi / M_{Pl}$. The see-saw now gives light Majorana masses: $m_\nu 
\approx (m_I v / M_{Pl})^2 / (m_I^2 /M_{Pl}) \approx v^2 / M_{Pl}$ --- 
the usual result!

Even simpler possibilities occur: the $G$ quantum numbers may prevent a 
right-handed Majorana mass to very high order, so that the dominant mass 
term is Dirac.  For example the operator
\begin{equation}
{1 \over M_{Pl}^{2}} [X^\dagger LNH_u]_D
\label{eq:xdaggerlnh}
\end{equation}
dominates either if the operator (\ref{eq:xlnh}) is forbidden by $G$ 
or if $X$ does not have an $A$ component vev, and 
gives light Dirac neutrinos of mass $m_\nu \approx F v / M_{Pl}
\approx v^2 / M_{Pl}$.


In section two, we study the general low-energy effective theory for the 
interactions of $L$ and $N$ with $H_u$ and $X$, with a view to studying 
the interesting forms for the neutrino mass matrices. The above
Majorana and Dirac cases are considered further. 

What are the consequences of our proposal that neutrino masses are
suppressed relative to charged lepton masses by supersymmetry
breaking factors? While the most immediate consequence is that it
opens up a new class of models for neutrino masses, the most important
consequence may be that supersymmetric phenomenology can be
drastically altered. This is largely due to the possibility that the
right-handed sneutrino, $\tilde{n}$, can now be at the weak scale and 
couple via a new $A$-type interaction:\footnote{If we wish to
  combine the unsuppressed $A$ terms arising from (\ref{eq:a}) with the light 
  Dirac neutrinos whose masses are generated by (\ref{eq:xdaggerlnh}), 
  we must require that the $A$ component vev of $X$ is small,
  $\a{X}\le v$.}
\begin{equation}
{ 1 \over M_{Pl}} [XLNH_u]_F \supset v \; \tilde{l} \tilde{n} h_u.
\label{eq:a}
\end{equation}
The structure of this interaction is investigated in section 3,
its consequences for the sneutrino mass spectrum is studied in section
4, and its consequences for the lightest sneutrino
as the cosmological dark matter in section 5. 
We find two interesting cases where
this occurs, and each case predicts a characteristic signal in
upcoming experiments to directly detect halo dark matter.

The $A$-term interaction $\tilde{l} \tilde{n} h_u$ can significantly
alter the decay branching ratios for the charged and neutral Higgs
bosons in supersymmetric theories. This is studied in section 6, where
we find that, for certain ranges of parameters, the decays $h
\rightarrow \tilde{\nu} \tilde{n}$ and $H^\pm \rightarrow
\tilde{l}^\pm \tilde{n}$ can be the dominant decay modes. Further
consequences for collider phenomenology, arising from the $A$-term
changing the decay chain of $\tilde{\nu}$, are discussed in section 7.
Finally, in section 8 we study the rare lepton flavor violation
implied by our mechanism for neutrino mass generation.

\section{Small neutrino masses from $F$ term SUSY breaking }
\label{sec:bigsec}

We begin by considering the low-energy effective theory which describes
the interactions of the leptons $L$ and $N$ with the Higgs doublet $H_u$
and the fields $X$ which spontaneously break both supersymmetry and the
global symmetry group $G$. We impose
$R$-parity, which changes the sign of the $L$ and $N$ superfields, as
well as the superspace coordinate $\theta$, but
leaves $H_u$ and $X$ unchanged. Expanding in powers of $1 / M$,
\begin{eqnarray}
\mathcal{L}_{eff} &=& \left[c_{4,1} X NN + c_{4,2} LNH_u \right]_F \\ \nonumber
&+&{1 \over M} \left( \left[{c_{5,1}} XLNH_u 
        + {c_{5,2}} (LH_u)^{2}+{c_{5,3}}N^4 + 
        {c_{5,4}} (XN)^2\right]_F + \left[{c_{5,5} 
        } X^{\dagger} NN\right]_D \right) \\ \nonumber
&+&{1 \over M^2} \left( \left[{c_{6,1} } X^{\dagger} LNH_u 
        + {c_{6,2}} X X^\dagger NN  \right]_D+\left[{c_{6,3} }X^{3}
    N^{2}\right]_F + \ldots \right)
        \label{eqnarray:operators}
\end{eqnarray}
Here and below, the energy scale $M$ is the ultraviolet cutoff of the 
low-energy theory, such as the Planck scale or the GUT-scale.

We have included all possible operators even though many may be excluded
by the global symmetry $G$, depending on the model. The bare mass $[NN]_F$
is always forbidden by $G$ and is not shown. Likewise, the mass terms
$XN$ are forbidden by $R$-parity. 
The flavor structure is not shown explicitly --- there are three $L$
fields and one or more $N$ field, and in general the coefficients
$c_{i,j}$ are power series in flavor symmetry breaking parameters $\phi /
M$.
There are other dimension 
six operators, such as $\left[ N^{\dagger} N L^{\dagger} 
L \right]_D$, but these will not affect the structure of the model nor the 
phenomenology, so we will not discuss them. It is possible that 
lepton-number violating dimension seven 
operators have important consequences, and we will discuss some 
of the consequences within the context of dark matter.

Before we consider particular symmetries, let us explore which
combinations of operators are phenomenologically interesting. In later 
sections, we will exhibit particular symmetries which realize these
scenarios. 
Since $N$ must appear in combination with some of the $X_{i}$ 
superfields, $c_{4,2}=c_{5,3}=0$. If we allow $c_{4,1}\ne 
0$, when the $X_{i}$ get $F$ 
component vevs, this term will generate an intermediate scale 
mass for the $N$ scalar, $\tilde n$, and this case is thus of less 
phenomenological interest. 
Furthermore, precisely the same 
term must be omitted in the MSSM ($ X H_{u} H_{d}$), and its omission 
here seems very natural. For these reasons we will take $c_{4,1}=0$.%
\footnote{If the $A$-component vev of $X$ is zero, 
one might also think that $c_{4,1}$ must be zero for another 
reason. If one allowed such a large supersymmetry breaking mass for
$N$, such that the fermion was present in the weak-scale theory, but
the scalar was not,  
one might worry that loop effects would destabilize the 
$v/M$ hierarchy. However, all dangerous 
diagrams that appear are suppressed by small Yukawa couplings 
and are harmless.}

The structure of the theory can vary, depending on a few elements.
In particular, if $X$ develops an $A$ component vev, 
the size of the Yukawa couplings will be different. There are three different 
natural values for the $A$ component: $M$, $\sqrt{F}$ and zero. 
If it is $M$, then the Yukawa couplings from the $c_{5,1}$ term in 
(10) are 
order one, which is phenomenologically unacceptable. We consider the 
other two alternatives in greater detail. We will begin by considering 
situations with only one generation.

\subsection{One Light Dirac Neutrino}
\label{sec:sdirac}
If the $A$ component of $X$ is zero, but $X$ does gain an $F$ component, 
we generate Yukawas
\begin{equation}
      \left[{X^{\dagger} \over M}  LH_u N\right]_D
= \frac{F}{M}\left[ LH_uN\right]_F.
\label{eq:yukawas1}
\end{equation}

If we assume that $G_{F}$ sets $c_{5,5}=0$ in (10), so 
that there is no Majorana mass for the right-handed neutrino, then we have generated
a Yukawa of the order $M_{SUSY}/M$. When the Higgs field takes on 
a vev, we then have a mass for the neutrino $O(v^{2}/M)$.
This is astonishing, because we now have a naturally light {\em Dirac} 
neutrino, with a mass of the correct size to explain the observed 
phenomena associated with neutrino mass\footnote{Whether these are precisely 
the right size is not of particular concern. This is an effective theory and 
$M$ could easily be $M_{GUT}$, or some other scale, in which case the Yukawas are 
larger.}. If this is correct, 
then experiments studying neutrino mass have {\em already} begun to 
probe the structure of supersymmetry breaking!

A small Yukawa coupling can, in principle, at least, be understood
with the Froggatt-Nielsen mechanism \cite{FN}, through the ratio $v/M$
of the vev of a supersymmetry conserving spurion over some higher mass scale.
Such a vev could arise from dimensional transmutation, or, as recently
discussed by \cite{cvetic}, radiative symmetry breaking, which was
employed by \cite{langacker} to achieve such light sterile
states. The possibility of using SUSY breaking operators to
generate light sterile states was initially explored by \cite{dvali}.
However, in contrast to these theories, the present class of models features a natural
mechanism for generating weak scale $A$ terms through the operator\footnote{In fact, if
  $\vev{X}=\theta^2 F$ is generated in the global SUSY limit,
  supergravity effects will generate a small shift in the $A$
  component, giving $\vev{X}\sim F/M_{Pl}+\theta^2 F$ \cite{izawa}.
  Thus, the large $A$ terms and the small Dirac neutrino masses
   can be generated from the operator of
  (\ref{eq:Aterms1}) alone \cite{babu, nomura}.}
\begin{equation}
      \f{ {X \over M}  LH_u N} 
= \a{ { F \over M} LH_uN}. 
\label{eq:Aterms1}
\end{equation}

Operators like (\ref{eq:yukawas1}) and (\ref{eq:Aterms1}) could be selected 
by a symmetry $U(1)_N\otimes U(1)_L$, with 
fields $X$ and $\overline X$, with charges $(1,1)$, and $(-1,-1)$,
respectively. $E$, $L$ and $N$ have charges $(-1,0)$, $(1,0)$ and
$(0,1)$, respectively. 
With these charges, the only operators of dimension six or less allowed are
\begin{equation}
\d{ 
{c_{5,1} \over M} X^\dagger LNH_u} +\f{
{c_{6,1} \over M^2} {\overline X} LNH}.
\label{eq:sdex}
\end{equation}
The superfields $X$ and $\overline X$ could acquire $F$ component vevs, but no $A$ 
component vevs if embedded in an O'Raifeartaigh model. Given a
superpotential
\begin{equation}
W = S (Y \overline Y - \mu^2) + Y^2 \overline X + \overline Y^2 X,
\label{eq:orsp}
\end{equation}
the minimum of the scalar potential will occur with $\vev{y}=\vev{\overline
y}=\mu/\sqrt{3}$. Here $Y$ and $\overline Y$ have charges $(1/2,1/2)$
and $(-1/2,-1/2)$, respectively. 
Two linear combinations of $s$, $x$ and $\overline
x$ will have positive masses at tree level, while the third
independent combination will get its mass in the one-loop effective
potential, stabilizing $\vev{x}=\vev{{\overline x}}=0$.
Note that the presence of the superpotential term 
$\left[{\overline Y^2\over M}
LNH_u\right]_F$ generates a contribution to the Yukawas of the same order of
magnitude as that from $\left[{ X^\dagger \over M} LNH_u\right]_D$. 
Note also that this breaks $U(1)_N\otimes U(1)_L$, but
preserves $U(1)_{L-N}$, which is the ordinary lepton number symmetry.

We will refer to this scenario, in which right-handed neutrinos couple
with suppressed Yukawas, but have no Majorana masses, 
as ``Dirac masses from supersymmetry breaking,'' or ``sDirac'' for
short.

\subsection{One Light Majorana Neutrino}
\label{sec:smajorana}
An alternative to generating light Dirac neutrino masses from
supersymmetry breaking is to instead generate light Majorana
masses. We begin by considering the operators
\begin{equation}
\d{ {X^\dagger \over M} NN} + \f{{X\over M} LNH_u}.
\label{eq:smajops}
\end{equation}
Such terms could be justified by an R symmetry, where N has R charge
$2/3$, $L$ and $H$ both have R charge $0$, and $X$ has R charge $4/3$.
If $X$ takes on an $A$ component vev
$\vev{X}|_{\theta=0}=\sqrt{F}$, as well as an $F$ component vev, $F\approx 
(10^{11}\gev)^2\approx v \mpl$\footnote{For instance, in the SUSY
  breaking model of \cite{yanagida}, a chiral superfield naturally
  develops $A$ and $F$ components of the same order of magnitude under
  the dynamical assumption that a constant appearing in the Kahler
  potential is negative \cite{babu}.}, then the second term in (\ref{eq:smajops}) 
generates a weak scale $A$ term, but now generates a Yukawa coupling to 
the Higgs as well roughly of the size 
$\sqrt{F}/M \approx 10^{-8}$. If this were the end of the story,
then we would have a 
(Dirac) neutrino mass $\sim 1 {\rm keV}$. However, the first term of
(\ref{eq:smajops}) will now generate a Majorana mass for $N$ of the
order $F/M \approx 100\gev$, yielding a LR neutrino mass matrix 
\begin{equation}
        \pmatrix{ 0 & v \sqrt{F}/M \cr v \sqrt{F}/M & F/M}.
        \label{eq:LRmatrix}
\end{equation}
After integrating out the heavy $N$ fermion, we are left with a 
Majorana mass for the neutrino with a size $m_{\nu} \approx {v^{2} 
 \over M }$, again reproducing the see-saw result.%
\footnote{
A weak scale Majorana mass for $N$ could have been generated with
nonzero $c_{5,4}$ as well. However, the scalars would then  
have a supersymmetry breaking mass squared $O(\sqrt{F} M_{W})$. Thus, 
for the same reasons we took $c_{4,1}=0$, we do not consider the case 
where $G_{F}$ allows nonzero $c_{5,4}$.}


We will refer to this scenario, in which the right-handed neutrinos
have Yukawas $O(\sqrt{F}/M)$ and weak scale Majorana masses, as
``Majorana mass from supersymmetry breaking'', or sMajorana for short.

\section{Flavor Structures}
\label{sec:flavstruc}

In section \ref{sec:bigsec}, we concerned ourselves simply with the
origin of neutrino mass itself, but did not address the additional
question of what determines the structures of these masses when we
include additional generations. As was discussed in the introduction,
we are considering a scenario in which the global symmetry of the
theory is $G_F \otimes G\otimes SUSY$. $G$ must include some symmetry to keep 
the Higgs doublets and right-handed neutrinos light, 
and $G_F$ may include symmetries such as
$U(3)^6$ which relate the different generations. 

The key feature of our model is that the supersymmetry breaking
spurions also contain charges under $G$. When these spurions acquire 
$F$, and possibly $A$ component vevs, they break $G$. Of course,
they need not be charged merely under $G$, but potentially under some
larger group $H$, where $G_F \otimes G \supset H \supset G$. 
In the most minimal framework, $H=G$ and would contain
only those symmetries which are necessary to suppress the $\mu$ term 
and the right-handed neutrino masses, for instance $U(1)_L \otimes
U(1)_N$ in section \ref{sec:sdirac}, or the $R$ symmetry in section
\ref{sec:smajorana}. 

With such an assumption, the textures of the neutrino mass matrices
and the $A$ terms would be determined by
supersymmetry preserving elements of the theory. For instance, in the
sDirac scenario, the couplings would be given by
\begin{eqnarray}
   \f{\lambda_{ij} H_u L^i N^j} &=& 
  \d{X^\dagger  \Lambda_{ij} H_u L^i N^j}, \\
  A_{ij} h_u \tilde l^i \tilde n^j &=& \f{ {\overline X}
  \Lambda'_{ij} H_u L^i N^j},
\end{eqnarray}
where $\Lambda_{ij}$ and $\Lambda_{ij}'$ are 
supersymmetry {\em preserving} but flavor
breaking spurions. As such, an explanation of structure of the
Yukawas and $A$ terms is beyond the scope of our scenarios. However,
because of this, in the presence of a flavor symmetry, we are able to
easily relate the structure of $A_{ij}$ and $\lambda_{ij}$. 

However, in the sMajorana case, the couplings will be given by
\begin{eqnarray}
A_{ij} h_u \tilde l^i \tilde n^j &=& 
 \f{X_F \Lambda_{ij} H_u L^i N^j},\\
 \f{\lambda_{ij} H_u L^i N^j} &=& 
 \f{X_A \Lambda_{ij} H_u L^i N^j}.
\end{eqnarray}
Here, the Yukawas are precisely the same as the $A$ terms, but due to
the potential mixing of the $N$'s, it is impossible to necessarily
relate the $A$ term matrix to the neutrino mass matrix obtained by the
see-saw mechanism.

We consider this to be the minimal scenario, in which $H$ is as 
small as possible, so to speak. However, $H$ can easily be much
larger. Indeed, people have investigated the possible effects of baryon and
lepton number violation operators from supersymmetry breaking \cite{blvio}. 
A priori, there is no reason why we should reject the
possibility that $H=G_F\otimes G$. If that were the case, we would write the
sDirac couplings as 
\begin{eqnarray}
  \f{\lambda_{ij} H_u L^i N^j} &=& \d{
  X^\dagger_{ij} H_u L^i N^j}, \\
  A_{ij} h_u \tilde l^i \tilde n^j &=& \f{{\overline
    X}_{ij} H_u L^i N^j}. 
\end{eqnarray}
Now the spurions $X$ and $\overline X$ carry generation indices
themselves! We assume that charged fermion Yukawas are generated
in a supersymmetry preserving sector of the theory. 
Since the flavor structure of $X$ and $\overline X$ is
entirely determined in the supersymmetry breaking sector, they need not 
be aligned with those of the supersymmetry preserving
Yukawas. Consequently, a large mixing between $\nu_\mu$ and $\nu_\tau$ 
is natural. Of course, the small mixing between $\nu_e$
and this heavy state ($\theta_{e3}<0.16$ as required by CHOOZ
\cite{CHOOZ}), must be viewed as somewhat of an accident, but not
necessarily a fine-tuning. This is similar to the anarchy proposal of
\cite{anarchy}, except that here we need not relate the Yukawas of 
$\nu$ and $e$, and a hierarchy of eigenvalues could still
possibly occur in $X$. Consequently, even for sDirac neutrinos,
anarchic aspects of the theory are reasonable.
Whether this is compatible with supersymmetric flavor changing
constraints is an important question. We will address this
further in section \ref{sec:lfv}.

It is important to note that we do not need three right-handed
neutrinos for the cases of sections \ref{sec:sdirac} and
\ref{sec:smajorana}. The presence of just two $N$ states is enough to
generate either two massive Dirac or two massive Majorana
neutrinos. The remaining neutrino is simply a massless Weyl
neutrino. In a certain sense, this is more minimal than with three
$N$'s, but the phenomenology is largely unchanged.

One limit of this could be that there is, in fact, only one $N$. Here, 
one might generate Majorana masses for the neutrinos through an
ordinary $GUT$ see-saw, while a sDirac or sSterile $N$ would contribute a 
fourth mass eigenstate, resulting in four Majorana neutrinos. 
Given appropriate $G_F$ charges, other, more exotic
possibilities may exist, such as combinations of sDirac and sMajorana
neutrinos.

\section{The Sneutrino Mass Matrix}\label{sec:mm}
In the MSSM, the sneutrino and charged slepton masses are intimately 
related:
\begin{equation}
        m_{\tilde{\nu}}^{2}=m_L^{2} +{1 \over 2} m_{Z}^{2} \cos2 
        \beta, \hspace{.4 in} 
        m_{\tilde{l}_L}^{2}=m_L^{2}+\left(\sin^{2}\theta_{W}-{1 \over 
        2}\right)
        m_{Z}^{2} \cos 2 \beta,
\end{equation}
where $m_{L}$ is the soft scalar mass for the left-handed 
sleptons.  For $\tan \beta > 1$, $\cos 2\beta<0$ and the $D$-term 
splitting pushes the sneutrino mass down and the charged slepton mass 
up. The present experimental bound $m_{\tilde{l}_{L}}>70$GeV 
still allows
for light sneutrinos due to this splitting.  However, if 
in the future it becomes established that $m_{\tilde{l}_{L}}$ is very 
large, 
much of the phenomenology associated with light 
sneutrinos will be ruled out in the MSSM.  In our 
model, with light right-handed sneutrinos, the story is quite
different, both because
the $A$ terms provide an additional source of splitting between
the sneutrino and charged slepton masses, and because the right-handed
sneutrino mass is not linked to slepton masses by gauge invariance. 
Thus, even if $m_{L}$ is quite large it is still possible
to have significant change in phenomenology that would otherwise be absent.

\begin{figure}[tbp]
   \begin{center}
     \leavevmode
     \psfig{file=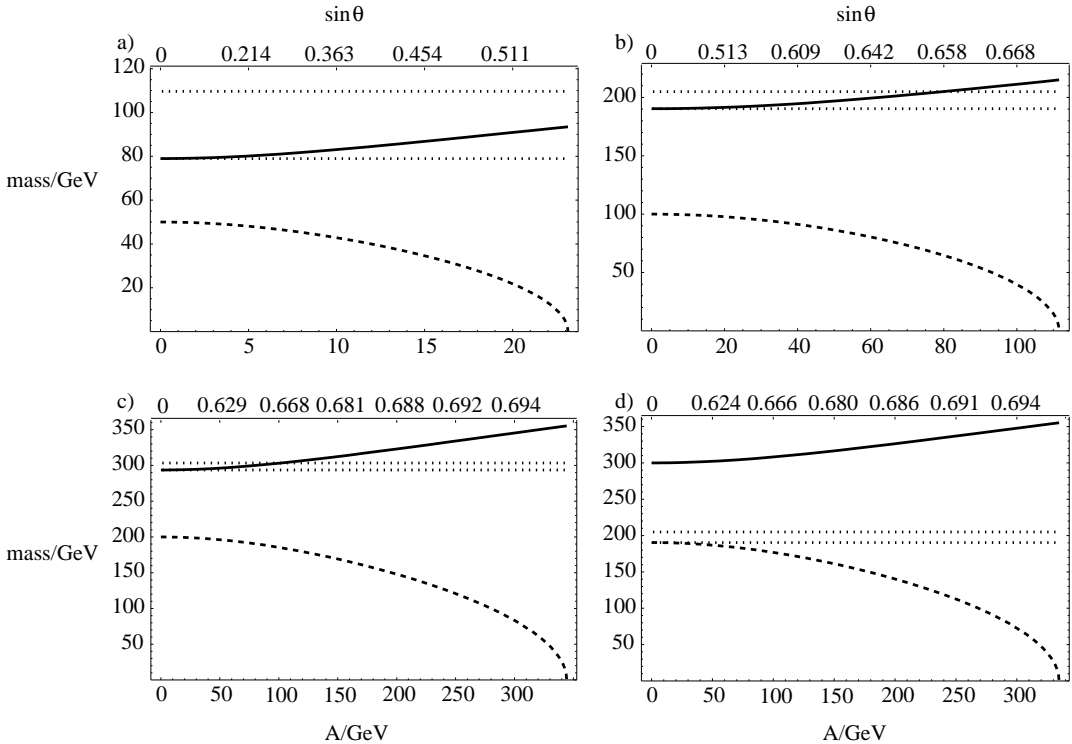,width=1.0\textwidth}
     \caption{Slepton mass spectra as a function of $A$ for (a) $m_{L} = 
       100{\rm GeV}$, $m_{R} = 50{\rm GeV}$, (b) $m_{L} = 
       200{\rm GeV}$, $m_{R} = 100{\rm GeV}$, (c) $m_{L} = 
       300{\rm GeV}$, $m_{R} = 200{\rm GeV}$, and (d) $m_{L} = 
       200{\rm GeV}$, $m_{R} = 300{\rm GeV}$. The solid line is
       the mass of the heavier sneutrino, the dashed line that of the
       lighter sneutrino. The dotted lines are the masses of the
       sneutrino (lower dotted) and charged slepton (higher dotted) in the 
       MSSM. Curves are drawn for $\tan \beta = 5$, and are 
       relatively insensitive to $\tan \beta$.}
     \label{fig:spectra}
   \end{center}
\end{figure}

To better understand the spectrum, we 
consider a single sneutrino generation with mass-squared matrix
\begin{equation}
        m_{\tilde{\nu}}^{2}=\pmatrix{{m}^{2}_{L} +{1 \over 2} m_{Z}^{2} \cos2 
        \beta & A v \sin\beta \cr A v \sin\beta
        &{m}^{2}_{R}  }.
\label{eq:massmatrix}
\end{equation}
Given that $m_{L}$, $m_{R}$ and $A$ are independent parameters, 
this matrix can have very different eigenvalues. We plot the mass spectra for 
various choices of $m_{L}$ and $m_{R}$ as a function of $A$ in 
figure \ref{fig:spectra}.

An independent lower bound on the sneutrino mass in the MSSM, 
$m_{\tilde{\nu}}>44$ GeV,  
comes from the measurement of the invisible with of the $Z$, and is also 
altered by the addition of right-handed sneutrinos.  The lightest 
sneutrino in our model is a superposition of left and right-handed 
states:
\begin{equation}
        \tilde{\nu}_{1}= -\tilde{\nu}_{L} \sin\theta+
 \tilde{\nu}_{R}\cos\theta.
\end{equation}
If this state is light enough to be produced in $Z$ decays, its 
contribution to the $Z$ width is given by
\begin{equation}
        \delta \Gamma = {\sin^4\theta \over 2} \left( 1 - \left({2 m_{\tilde \nu_{1}} \over 
        m_{Z}}\right)^{2}\right)^{3/2} \Gamma_{\nu}
        \label{eq:invwidth}
\end{equation}
where $\Gamma_{\nu}=167{\rm MeV}$ is the $Z$ width to ordinary neutrinos. 
If we take the current LEP limit of $2{\rm
  MeV}$ \cite{zneutrinolimits}, the $\sin^4 
\theta$ factor allows us to evade the bounds regardless of mass 
provided $\sin\theta<0.39$, which is a very mild constraint. 

\begin{figure}[tbp]
   \begin{center}
     \leavevmode
     \psfig{file=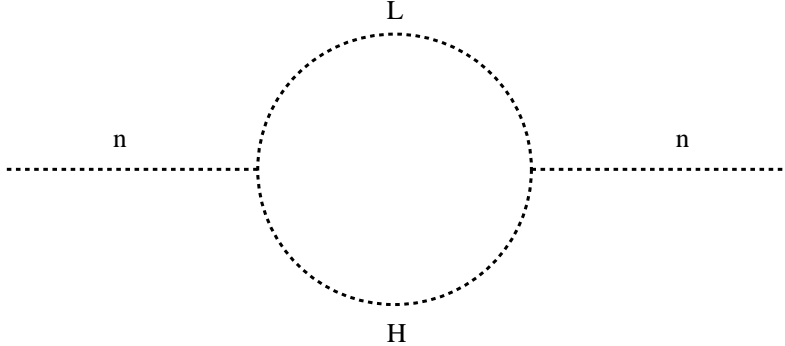,width=0.65\textwidth}
     \caption{Loops contributing to the running of $m_{\tilde n}$.}
     \label{fig:aloop}
   \end{center}
\end{figure}

Although $m_{L}$, $m_{R}$ and $A$ are independent parameters, we can 
gain some intuition for their sizes from their renormalization group 
running. In fact, it is somewhat natural to have $m_{R}<m_{L}$ in 
the low-energy theory. 
The running of $m_L$ and $m_R$ is governed by
\begin{equation}
\frac{ d m_{L}^2}{d t} = -{3 \over 16 \pi^2} g^2 M_2^2 - {3
  \over 80 \pi^2} g_Y^2 M_1^2 + {1\over 16 \pi^2} A^2,
\label{eq:nurunning}
\end{equation}
\begin{equation}
{ d m_{R}^2 \over d t} = {2\over 16 \pi^2} A^2,
\label{eq:nrunning}
\end{equation}
where $t = {\rm ln}(\mu^2/\mu_0^2)$.
Since $\tilde n$ is a standard model singlet, there are no gaugino
loops to drive its mass upward as we run the energy scale down from
$M_{Pl}$ to $M_{W}$. Likewise, there are new, sizeable 
 loop diagrams arising from the 
$A$ terms (figure \ref{fig:aloop}), which push 
the soft masses down. However, two states ($\tilde l$ and $\tilde \nu$) can 
propagate in the loop contributing to $m_{R}$, while only one ($\tilde 
n$) can propagate in the loop contributing to $m_{L}$, 
pushing $m_{\tilde n}$ down faster than $m_{\tilde \nu}$.

In summary, the mass matrix can have 
two {\em very} different mass eigenstates, and there can be 
particles that couple very much like $\tilde \nu$, but with 
suppressed couplings and masses unrelated to our expectations from the 
MSSM. The lightest is likely to be dominantly $\tilde n$, such that 
its mass is not restricted by $Z$ decay data.

\section{Sneutrino Dark Matter}
One of the appealing features of  R-parity conserving 
supersymmetric theories with gravity-mediated supersymmetry breaking
is that the lightest superpartner (LSP) is a 
good candidate for dark matter. Searches for superheavy 
hydrogen have ruled out a charged LSP, leaving the 
neutralino and the sneutrino as candidates for dark matter.

A number of direct searches for dark matter have been carried out 
\cite{CDMS,DAMA,HMGe} which have essentially excluded sneutrino dark 
matter in the MSSM unless $m_{\tilde \nu} < 10$ GeV. However, as we 
have already discussed, measurements of the invisible width of the $Z$ exclude 
such a light sneutrino. Within our framework, the $Z$ width provides
only a mild constraint, and we are free to explore the possibility of a light 
sneutrino dark matter candidate. A second, equally important point is
that the $\sin^2\theta$ suppression of the light sneutrino coupling to
the $Z$ boson greatly reduces the sneutrino-nucleon cross section,
making it possible even for a heavier $\tilde{\nu}$ to evade direct
detection.
Finally, if we include lepton
number violation, the lightest sneutrino cannot scatter elastically
via $Z$ exchange \cite{hallmurayamamoroi}, further diminishing 
  the limits from direct searches. Sneutrino dark matter requires
  the presence of the $A$-term of (\ref{eq:a}), and hence is linked to 
  the other phenomenology of the interaction.
     
\subsection{Dark matter without lepton number violation}
\subsubsection{Light sneutrinos}
\begin{figure}
\begin{center}
     \leavevmode
     \psfig{file=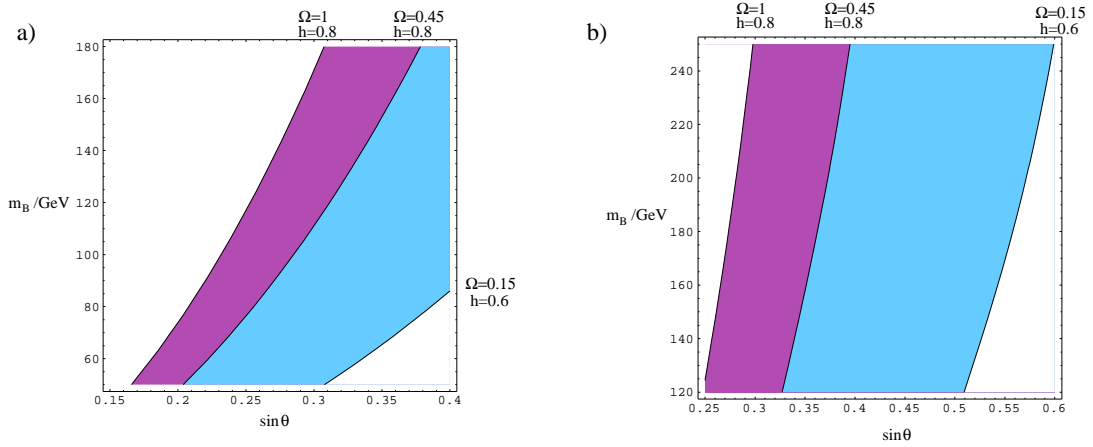,width=0.90\textwidth}
\caption{Contours of $\Omega h^{2}$, where $h$ is the normalized 
     Hubble parameter, as a function of the bino 
     mass $m_{B}$ (assuming GUT
   unification of gaugino masses) and $\sin \theta$. Both shaded
   regions yield relic densities below overclosure, with 
the lighter shaded region corresponding to values of $\Omega$
preferred by supernovae data \cite{perl}. In a) we take
   $m_{\tilde{\nu}}=10$ GeV, and in b) we take $m_{\tilde{\nu}}=100$
   GeV, $A=20$ GeV,
   $\tan\beta=50$, and $m_h=115$ GeV. For b), direct detection bounds are
   evaded only in the lepton-number-violating case.}
     \label{fig:relicdensity}
   \end{center}
 \end{figure}

We can determine the relic density of light sneutrinos
($m_{\tilde{\nu}}<10$ GeV) through standard 
methods \cite{susydm}. The dominant annihilation process is through
$t$-channel neutralino exchange. 
If the neutral wino exchange dominates, 
we find
\begin{equation}
        \Omega_{\tilde \nu_{1}}h^{2} \approx \left({ 
        M_{\tilde W} \over 100 {\rm GeV} } \right)^2 \left({ 0.19
        \over \sin\theta}\right)^4,
        \label{eq:omega}
\end{equation}
where $h$ is the normalized Hubble parameter. We find it highly
significant that for wino masses of roughly 
$100 {\rm GeV}$ and angles roughly a half of the limit from the
invisible $Z$ width, we have a cosmologically 
interesting amount of sneutrino dark matter. 
We show in figure \ref{fig:relicdensity}a the relic density of 
sneutrino dark matter considering all annihilation processes. If this
scenario is correct, and $\tilde \nu_{1}$ is the dark matter, 
then the mixing angle must be near the limit from the invisible $Z$ 
width measurements, making a future detection possible. In particular, 
such a sneutrino would almost certainly be seen in the upcoming CRESST 
experiment \cite{CRESST}.

Relic sneutrinos captured by the sun will annihilate into neutrinos
that can induce upward-going-muon events on earth \cite{krauss}.  The flux of these
muons is constrained to be less than $10^{-14}$ cm$^{-2}$
s$^{-1}$ \cite{indirect}.  For a 10 GeV sneutrino with
$\sin\theta=.2$,
we calculate a flux, using the formulae of \cite{susydm}, that is
roughly three times this, assuming that all neutrinos produced are
muon-type when they reach the earth\footnote{For light sneutrinos 
and large $A$, the rate for sneutrino capture
 by the sun is dominated by Higgs exchange and the
  flux can be much larger.  Here we assume $A \simeq 10$ GeV, so that the
  capture rate is dominated by $Z$ exchange.} (we find a much smaller flux
due to capture by the earth itself).   The actual muon flux could be
suppressed depending on the flavor of the LSP sneutrino and on neutrino
oscillation parameters; for instance, for an electron-type sneutrino
and the small angle MSW solution to the solar neutrino problem, the
flux would be smaller by roughly a factor of a thousand.

\begin{figure}
     \begin{center}
     \leavevmode
     \hspace{-.9cm}\psfig{file=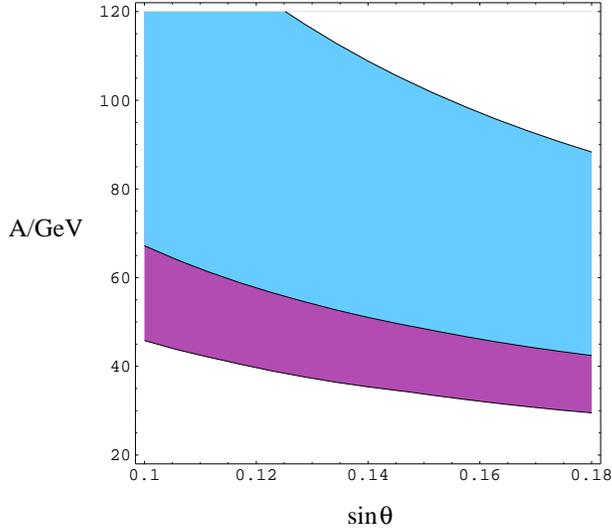,width=.5\textwidth}
\caption{Contours of $\Omega h^{2}$ for $m_{\tilde{\nu}}=100$ GeV as a function of 
$\sin\theta$ and $A$. The meanings
       of the lighter and darker shaded regions are the same as in
       figures 3a) and 3b).  
We take  $\tan\beta=50$, $m_h=115$ GeV, and a bino mass of
       $200$ GeV, and we assume GUT
   unification of gaugino masses.}
     \label{fig:relicdensity2}
   \end{center}
 \end{figure}

\subsubsection{Heavier sneutrinos}
For heavier sneutrinos,
the strongest current direct detection limit comes from CDMS
\cite{CDMS}, which, under the assumption of $A^2$
scaling, constrains the nucleon-relic cross section to be
less than $(2-3) \cdot 10^{-42}$ cm$^2$ for relic masses of $O(100$ GeV).
The cross section for ordinary sneutrino-nucleus scattering is
\begin{equation}
\sigma={G_F^2 \over 2 \pi} \mu^2 \left((A-Z)-(1-4 \sin^2 \theta_W) Z \right)^2,
\end{equation}
where $\mu$ is the sneutrino-nucleus reduced mass.
In our framework this cross section comes with an additional
$\sin^4\theta$ suppression, implying that for $m_{\tilde{\nu}}$ much
larger than the nucleon mass $m_N$, the CDMS constraint can be evaded by
requiring
\begin{equation}
\sin^4\theta<2 \pi \left( {A\over(A-Z)-(1-4 \sin^2 \theta_W) Z
    }\right)^2 \left({2 \cdot 10^{-42} {\rm cm^2}\over G_F^2 m_N^2}\right).
\end{equation}
Taking $A=73$ and $Z=32$ for Ge$^{73}$ gives $\sin \theta<.17$.  
The DAMA collaboration \cite{DAMA} has reported a
positive signal corresponding to a relic-nucleon cross section of
roughly $2-10 \cdot 10^{-42}$ cm$^2$ and a relic mass $\sim 30-100$
GeV
In our framework this range in cross section corresponds approximately
to $.17<\sin\theta<.25$. 


In figure \ref{fig:relicdensity2}, we plot contours of $\Omega h^2$ for
$m_{\tilde{\nu}}=100$ GeV and a bino mass of 200 GeV.
For this choice of parameters, and for large enough $A$,
the dominant annihilation processes in
the early universe are s-channel Higgs exchange into $W^+W^-$ and $Z$
pairs, which have cross sections proportional to $A^2 \sin^2 2\theta$ 
rather than $\sin^4\theta$.  These are also the dominant annihilation
processes for sneutrinos trapped in the sun.  This is important because the
alternative process is annihilation directly into neutrinos via t-channel
neutralino exchange, which would likely lead
to a much larger signal at indirect detection experiments.  Assuming
that $1/3$ of all neutrinos produced in the sun are muon-type upon
reaching the earth, we find that indirect detection constrains
$\sin\theta<.18$ for the parameters we have chosen, comparable to the
CDMS constraint.  The interesting relic abundances indicated in figure
\ref{fig:relicdensity2} lead us to conclude that the prospects for sneutrino
dark matter with $m_{\tilde{\nu}} \sim 100$ GeV are quite interesting
in our model.

\subsection{Dark matter with lepton number violation}
As previously explored \cite{hallmurayamamoroi}, 
the presence of lepton number violation 
changes the limits from direct searches for dark matter significantly. 
In our model, lepton number violation can reside in
the $m_{nn}^{2} \tilde n \tilde n$ term in the
Lagrangian. Such a term 
could easily arise from dimension seven operators in the
Lagrangian, such as 
\begin{equation}
\d{{X^\dagger X X^\dagger \over M^3} N^2}.
\label{eq:splitop}
\end{equation}

The presence of this lepton number violation splits the CP-even and odd 
states $\tilde \nu_{+}$ and $\tilde \nu_{-}$. However, the coupling to the $Z$ is 
off diagonal, i.e., $Z \tilde \nu_{+} \tilde \nu_{-}$. Consequently, 
for large enough $\Delta m = |m_{\tilde{\nu}_{+}}-m_{\tilde{\nu}_{-}}|$, 
the LSP sneutrino cannot scatter off nuclei via $Z$-exchange, 
eliminating constraints arising from 
CDMS 
, 
DAMA
, 
and the Heidleberg-Moscow Ge experiment \cite{HMGe}. More precisely, the
scattering is kinematically forbidden if $\Delta m>\beta_h^2 m_{\tilde
  \nu} m_A/2(m_{\tilde \nu} + m_A)$, where $m_A$ is the mass of the
target nucleus, and $\beta_h=10^{-3}$ for virialized halo particles on average.
For example, taking $m_{\tilde \nu}=100$ GeV and a Ge target, 
we require $\Delta m > 20\kev$. Since $\Delta m =
m^2_{nn}/m_{\tilde \nu}$, this corresponds to 
$m_{nn}^2 > (45\mev)^2$, which is of the order of what we
expect from (\ref{eq:splitop})\footnote{Somewhat higher values of
  $\Delta m$ may be required to prevent indirect detection due to
  sneutrino capture and annihilation in the sun.  Here we
  simply assume that $\Delta m$ is large enough to evade indirect
  detection as well.}.

The effects of lepton number violation in dark matter have 
been previously explored \cite{hallmurayamamoroi}. 
However, in the model previously proposed, there were no singlet
sneutrinos, so the mass splitting $\Delta m$ was
required to be 
adequately large so as to suppress coannihilation between $\tilde 
\nu_{+}$ and $\tilde \nu_{-}$ via $s$-channel $Z$ exchange. In our model, 
this process is further suppressed by $\sin^{4} 
\theta$ in the cross section, so that even with small mass splittings from 
dimension seven or higher operators, we can still generate a 
cosmologically interesting abundance.

Unlike \cite{hallmurayamamoroi}, we now have the $A \tilde 
\nu \tilde n h$ coupling, which yields an extra contribution 
to the scattering of the lightest sneutrino  
off of nuclei via Higgs exchange.  The coupling of the
Higgs to nucleons is larger than just that from scattering off of
valence quarks \cite{SV}, but it is still quite small.  
Using the numerical
value for the Higgs-nucleon coupling from \cite{tpc}, we find
that the sneutrino-nucleon cross section obtained from Higgs
exchange alone is\footnote{We take the decoupling limit for the Higgs
  sector.}
\begin{equation}
 \sigma =  
\left({A\sin\beta \sin
     2\theta - (\sqrt{2}M_Z^2/v) \cos 2\beta \sin^2 \theta  \over
              100\gev}\right)^{2} \left({ 100\gev \over  
        m_{\tilde \nu}}\right)^2 \left({115 \gev \over m_h}\right)^4
           \left(3 \times 10^{-43}\cm^{2}\right).
        \label{eq:csnuc}
\end{equation}
For broad ranges of parameters this cross section falls well below the
current direct detection limits.  For instance, it is quite reasonable
to consider  
%
$m_{\tilde
  \nu} \sim 100\gev$ even for values of $\sin\theta$ larger than those
that allow one to evade CDMS in the lepton-number conserving case.  
Future experiments \cite{genius,newcdms} should be able to probe an additional 
three orders of magnitude down from the present constraint, giving a
significant probe of sneutrino dark matter over a considerable
range of parameters, for both the lepton-number conserving and
lepton-number violating cases.

Since the dominant annihilation process in the early universe is
$s$-wave, there is little dependence of the relic abundance
on the sneutrino mass given that the
sneutrino is relatively light ($\lesssim 30$ GeV). For these
relatively light sneutrinos, figure \ref{fig:relicdensity}a is still qualitatively
applicable. For larger sneutrino masses, $Z$ and Higgs pole effects
or production of $W$ and $Z$ pairs can be relevant.
The sneutrino relic density is shown in figure \ref{fig:relicdensity}b for
$m_{\tilde{\nu}}=100$ GeV and $A=$ 20 GeV.  

The lepton number violating mass $m_{nn}^2$ can induce radiative
corrections to the neutrino mass through neutralino loops.
If the splitting $\Delta m$ is too large, the possibility exists of
generating neutrino masses radiatively which are large enough to 
affect the overall analysis. Such a possibility will be
explored elsewhere \cite{us}. For our purposes here, we limit
ourselves to the case where the mixing between $\tilde n$ and $\tilde
l$ is small enough to suppress this radiatively generated mass (which
is why a relatively small value for $A$ is taken in figure \ref{fig:relicdensity}b).

We conclude that the
possibility of evading direct detection through 
lepton number violation leads to another
interesting version of sneutrino dark matter in our framework. Moreover, the
elastic scattering of sneutrinos from nuclei via Higgs exchange
is just below the current limits, and potentially detectable at 
upcoming dark matter searches.

\section{Higgs Decays}
The unsuppressed $A \tilde l \tilde n h_{u}$ coupling in
our scenario can
lead to interesting collider phenomena.  If kinematically allowed, 
$\tilde{\nu}_{1} \tilde{\nu}_{1}^{*}$ is typically the 
dominant decay mode for the light Higgs.  There is a similar situation 
in the the MSSM \cite{djouadi}: provided the sneutrinos are sufficiently
light and that $\tan 
\beta$ is not too close to 1, the $\tilde{\nu} \tilde{\nu}^{*} 
h$ coupling proportional to $M_{Z} \cos 2\beta$ leads to a
partial width into sneutrinos 
that is larger than that into $b \overline{b}$ by 
two orders of magnitude.  Assuming that the sneutrinos decay invisibly
into $\chi^{0}_{1} \nu$ (or that the sneutrino itself is the LSP), 
a light Higgs that decays dominantly into sneutrinos would
be very difficult to discover at LHC, leaving the NLC the 
opportunity for discovery through the process $e^{+}e^{-}\rightarrow 
Z^{*}\rightarrow Zh$.  

In the MSSM, the $Z$ width measurement, the theoretical bound
$m_H \lesssim 130$ GeV, and the relation between the $\tilde{\nu}$ and
$\tilde{l}_L$
masses constrain the region of 
parameter space in which the light Higgs can  decay into sneutrinos.  For
example, if in the future it becomes established that $m_{{\tilde l}_L}
\gtrsim 105$ GeV, the decay $h\rightarrow \tilde{\nu}\tilde{\nu}^{*}$ 
will be ruled out in
the MSSM.

\begin{figure}
  \centerline{
\psfig{file=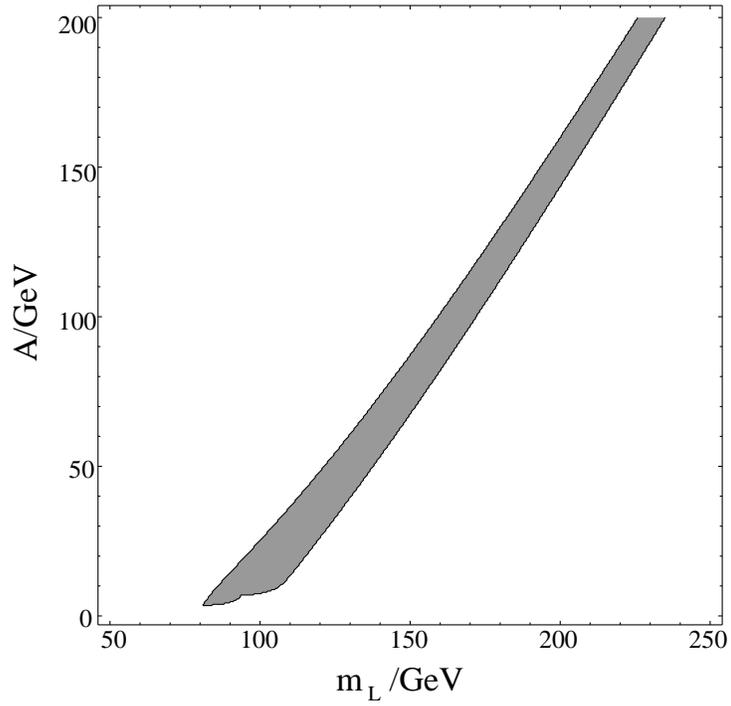,width=0.6\textwidth,angle=0}
}
 \caption{The region of parameter space in which the $Z$ width constraint
   is met and $\Gamma(h\rightarrow \tilde{\nu} 
\tilde{\nu}) > \Gamma(h \rightarrow b\overline{b})$ holds, for $\tan\beta=2$ and
$m_h=130$
GeV.  We take the splitting between $m_L$ and $m_R$ to be generated by 
RG running from the GUT scale, as discussed in the text.}  
\label{fig:hdecay}
\end{figure}

In the present scenario, however, the sneutrino mass spectrum is
expected to be quite different from that in the MSSM, as discussed in
section \ref{sec:mm}.  Even if $m_L$ (and therefore $m_{{\tilde l}_L}$) is quite
large, it is still possible for the light Higgs decay into sneutrinos to
be kinematically allowed.  To explore this possibility quantitatively, 
we consider a single generation of sneutrinos with the mass matrix of
equation (\ref{eq:massmatrix}), whose four free parameters are $m_L$,
$m_R$, $\tan\beta$, and $A$.  For simplicity we consider the case in which
the splitting
between $m_{L}^2$ and $m_{R}^2$ is generated by RG running from the
GUT scale to the weak scale, and adopt 
${m}^{2}_{R}={m}^{2}_{L}-.4 A^2 -.5 m_{1/2}^2$, with $m_{1/2}^2$, the
universal gaugino mass,
set to 100 GeV. The region of 
$({m}^{2}_{L}, A)$ parameter space in which the $Z$ width constraint is
met  and $\Gamma(h\rightarrow \tilde{\nu} 
\tilde{\nu}) > \Gamma(h \rightarrow b\overline{b})$ holds are displayed in
Figure \ref{fig:hdecay}
for $m_{h}=130$ GeV and $\tan\beta=2$ (the plot is very similar for
high $\tan\beta$).  As expected, 
there is a band in parameter space that yields invisible 
Higgs decays: for the region shown, the window for $A$ is roughly 10
GeV at fixed $m_L$.  The band persists for large $m_L$, with the
window for $A$ scaling as $\sim 1/ m_L$.  

The $A \tilde l \tilde n h_{u}$ coupling can also alter the details of
charged Higgs 
decays.  If $m_{H^{\pm}}<m_{t}$, one can look for the charged 
Higgs through the process $p{\overline p}\rightarrow t{\overline t}$,
with one or both of the top quarks decaying into $H^{+}b$ 
($H^{-}\overline{b}$).  The standard analysis exploits the fact that
the charged Higgs is coupled most strongly 
to $\tau\nu$ (in contrast to the universally coupled $W$), and so if
produced should lead to a surplus of $\tau$'s.
This analysis has been applied at the Tevatron to establish lower bounds on 
the charged Higgs mass for $\tan \beta \lesssim 1$ and $\tan\beta \gtrsim35$
 \cite{cdfdzero}.
The region of intermediate $\tan\beta$ will be only partially accessible
to Run II of the Tevatron, but should be covered entirely at the
LHC \cite{atlascms}.  Similarly, the Higgs search at LHC for high 
$\tan\beta$ employs the decay of heavy neutral Higgs states $H^{0}, 
A^{0} \rightarrow \tau^{+} \tau^{-}$, which can be suppressed due to 
the decay modes into sneutrinos. 

It has already been pointed out that if the charged 
Higgs decays into SUSY particles, the analysis 
changes drastically \cite{borzumati}.  If the 
decay into a charged slepton and a sneutrino is kinematically allowed, 
the MSSM lagrangian term $-(g/\sqrt{2}) 
M_{W} \sin 2\beta H^{+} \tilde{\nu}^{*} \tilde{l}_{L}$ ensures that
$H^{\pm}\rightarrow \tilde{\nu} 
\tilde{l}_{L}$ dominates over the Yukawa-coupling-induced 
decay into $\tau \nu$ for small $\tan\beta$.  Even for large  $\tan\beta$,
it is still possible for the decay into $\tilde{\nu}\tilde{\tau}$ 
to dominate due to the coupling 
$-(g/\sqrt{2})(m_{\tau}/M_{W})(\mu +A_{\tau} \tan\beta) 
H^{+}\tilde{\nu}^{*} \tilde{\tau}_{R}$ (of course, $\mu$ and $A_{\tau}$
must not take on values that push the lightest charged slepton mass
below the experimental bound).  In this case, the 
excess $\tau$'s produced will have lower energy than when 
produced directly via $H^{\pm}\rightarrow \tau^{+}\nu$.

An unsuppressed $A \tilde l \tilde n h_{u}$ coupling introduces the added 
possibility of $H^{\pm}\rightarrow \tilde{l}_{L}\tilde{n}$ decays.  
If kinematically allowed, this is another process that can dominate over the
direct decay
into fermions for small $\tan\beta$.  Once again, kinematical 
considerations for this decay are modified from the MSSM 
decays both because of the additional mass splitting between the sneutrinos
and charged sleptons, and because the $Z$-width constraint does not apply
to a 
sneutrino mass eigenstate that is chiefly right-handed. Thus it is
conceivable that $H^{\pm}\rightarrow \tilde{l}_{L}\tilde{n}$ is the
only SUSY decay mode allowed. Another 
important difference is that 
if one supposes that the flavor structure of the $A$ coupling is 
similar to that of the neutrino masses, then one expects the
charged Higgs to decay into $\tilde{\mu}\tilde{n}$ and 
$\tilde{\tau}\tilde{n}$ with similar probabilities, leading to an 
excess of both $\mu$'s and $\tau$'s over $e$'s.\footnote{Here we assume 
  that one neutrino mass is hierarchically heavier than the others,
  and that $H^{\pm}\rightarrow \tilde{l} \tilde{n}$ is kinematically 
  allowed for all flavors.}  

Even if $m_{H^{\pm}}>m_{t}$, it is still possible for $H^{\pm}
\rightarrow \tilde{l}_{L}\tilde{n}$ to be a  
dominant decay, for small to intermediate $\tan\beta$.  
The width is proportional to $(A 
\cos\beta)^2/m_H$, to be compared with 
$g^2 m_H/m_W^2 [(m_{t}\cot \beta)^2+(m_{b}\tan 
\beta)^2]$ for $H^{+}\rightarrow 
t\overline{b}$.  To obtain a competitive rate requires $A/m_H$ to be
sizable, which is most easily achieved kinematically when 
$m_H$ itself is large.
In this regime, decays into other SUSY particles are also likely 
to be important.

\section{Other Collider Phenomenology from $A$ Terms}
The unsuppressed $A$ terms can have other interesting consequences for collider
phenomenology, both due to their effect on the particle spectrum,
and because of the trilinear scalar vertex itself. 
Here we briefly consider a few examples, first for a visibly
decaying $\tilde{\nu}$, and second for an invisible $\tilde{\nu}$.  
We have already seen that it is natural in our scenario to have large mass
splittings among the various sneutrino states, so it is easily
conceivable that there will be sneutrinos in both categories.  
 
\subsection{Visibly decaying sneutrinos}

As in the standard framework of the MSSM, 
sneutrino decays into $\chi_2^0 \nu$ and $\chi_1^{\pm}
l^{\mp}$, if kinematically allowed, lead
to final states with, e.g., $2l{\not\!\!E}_T$, $ljj{\not\!\!E}_T$, or
$jj{\not\!\!E}_T$.  For example, possible decay chains
include\footnote{For now we ignore the role an additional, lighter 
$\tilde{\nu}$ might play in these decay chains.  For instance,
$\chi_2^0$ might decay invisibly into $\tilde{\nu}\nu$, as discussed
below.}
$\chi_2^0 \rightarrow \chi_1^0 Z^{(*)}/l \tilde{l}^{(*)}$ and
$\chi^{\pm}_1\rightarrow \chi_1^0 {W^\pm}^{(*)}/\nu \tilde{l}^{(*)}$,
followed by $Z^{(*)}\rightarrow ll/jj$,   ${W^\pm}^{(*)}
\rightarrow l\nu/jj$, and $\tilde{l}^{(*)}\rightarrow l\chi_1^0$.
A possible signal for sneutrino pair production at the NLC is
thus $4l{\not\!\!E}_T$.  Sneutrino pair production 
should be distinguishable from
neutralino pair production due to the different angular distributions
and the different $\beta$ dependences at threshold.  
As far as this signal is concerned, the distinctive feature of
our model is the admixture of the gauge
singlet $\tilde{n}$ in the sneutrino mass eigenstate.  Decomposing
the mass eigenstate as $\tilde{\nu} \sin\theta + \tilde{n}
\cos\theta$, the sneutrino pair production rate will be suppressed
by a factor $\sin^4\theta$ relative to its MSSM value.  By
performing a scan in energy one would be able to see the
$4l{\not\!\!E}_T$ signal
turn on for an isolated sneutrino mass eigenstate.  Then, knowing the
masses and mixings of the charginos and neutralinos, one could infer
from the measured rate the value of $\sin\theta$, and demonstrate
that the sneutrino produced is only partly left handed.
        
If a heavier $\tilde{\nu}_2$ state is sufficiently split from a lighter 
$\tilde{\nu}_1$, the unsuppressed $A$ term induces the decay 
$\tilde{\nu_2} \rightarrow \tilde{\nu_1} h$, providing
an interesting new way to produce Higgs
particles.  One might wonder how efficient this method of producing
Higgs would be at the LHC, through cascade decays such as 
$\tilde{q}\rightarrow q\chi^{\pm}$, $\chi^{\pm}\rightarrow l
\tilde{\nu}_2$, 
$\tilde{\nu_2} \rightarrow \tilde{\nu_1} h$.  The production cross
section of squarks and gluinos at the LHC depends sensitively on their 
masses.  Taking $m_{\tilde{q}} = 1.2 m_{\tilde{g}} =300$ GeV, the
cross section is $\sim 2$ nb at $\sqrt{s}=14$ TeV \cite{beenaker}, roughly fifty times
larger than the cross
section for gluon fusion Higgs production at that energy for $m_h \sim 
100-130$ GeV in the 
decoupling limit \cite{kunszt}, the regime we consider here.
If $m_{\tilde{q}} = 1.2 m_{\tilde{g}} =700$ GeV,
the cross sections are comparable.

Assuming that $A$ is sizable and that
$m_{\tilde{\nu}_2}-m_{\tilde{\nu}_1}>m_h$, the question of whether or
not there is an appreciable branching fraction for the cascades to
produce $\tilde{\nu}_1 h$ depends largely on $m_{\tilde{\nu}_2}$.  For
simplicity consider the case of a bino-like $\chi_1^0$ and
wino-like $\chi_2^0$ and
$\chi_1^{\pm}$, and take $\tilde{\nu}_2$ and $\tilde{\nu}_1$ to be
essentially left- and right-handed, respectively.  If
$m_{\tilde{\nu}_2}>m_{\chi^0_2},m_{\chi_1^{\pm}}$,
then $\tilde{\nu}_2$ will never be produced in the cascades, because
the colored particles decay via
$(\tilde{g}\rightarrow)\tilde{q}\rightarrow \chi_1^0 q/ \chi_2^0
q/\chi_1^{\pm}q$.  If
$m_{\tilde{\nu}_2}<m_{\chi^0_2},m_{\chi_1^{\pm}}$, then the branching
fraction for producing $\tilde{\nu}_1 h$ is the product of three
probabilities:  first, the probability of the gluino/squark decaying
into a neutralino or chargino heavier than $\tilde{\nu}_2$; second, the
probability of that gaugino decaying into
$\tilde{\nu}_2\nu/\tilde{\nu}_2 l$ rather than  into a lighter gaugino 
or $\tilde{l}\nu/\tilde{l} l$; and third, the probability that
$\tilde{\nu}_2$ decays into $\tilde{\nu}_1 h$ rather than
$\chi_1^0 \nu$.\footnote{Both BR($\chi\rightarrow\tilde{\nu}_1
  \nu/\tilde{\nu}_1 l$) and BR($\tilde{\nu}_2\rightarrow
  Z\tilde{\nu}_1$) are suppressed if $\tilde{\nu}_1$ is essentially
  right-handed.}

None of these probabilities is likely to be smaller than $\sim 1/$few
if $m_{\tilde{\nu}_2}<m_{\chi^0_2},m_{\chi_1^{\pm}}$, so in this case
the rate for Higgs production from $\tilde{\nu}_2$ decay at the
LHC could easily be comparable to or even much larger than that due to
$gg \rightarrow \gamma$.  Moreover, the cascade products (additional
jets, and an energetic lepton from $\chi_1^{\pm}\rightarrow
\tilde{\nu}_2 l$ decay, for instance), allow for detection via th
$h\rightarrow b {\overline b}$ mode, so that the signal is further enhanced
relative to $gg\rightarrow h \rightarrow \gamma \gamma$ by a factor
of a thousand\footnote{Thanks to Ian Hinchcliffe for pointing this out.}.

For example, suppose that
$m_{\chi_1^0}<m_{\tilde{\nu}_1}<m_{\tilde{\nu}_2}<m_{\tilde{l}}<m_{\chi_2^0},m_{\chi_1^{\pm}}<m_{\tilde{q}}= 
1.2 m_{\tilde{g}} 
= 300$ GeV.\footnote{If gaugino mass unification is imposed for this 
  mass ordering, then $m_{\tilde{g}}$ is forced to be much heavier,
  $>700$ GeV, in order for
$m_h<m_{\tilde{\nu}_2}<m_{\chi_2^0},m_{\chi_1^{\pm}}$
    to be satisfied.  In this case the squark and gluino production 
    cross section is comparable, at best, to the $gg\rightarrow h$ cross 
    section.}  Then the first probability is $\sim 1/2$, because
  $\tilde{q}_R$ couples to the bino, while $\tilde{q}_L$ prefers
  winos. The second probability is also $\sim 1/2$, because $\chi_0^2$ 
  and $\chi_1^{\pm}$ are equally likely to produce $\tilde{l}$ and
  $\tilde{\nu}_2$ and do not couple to $\chi_1^0$ in the limit that it 
  is pure bino.  The third probability is determined by
$\Gamma(\tilde{\nu}_2\rightarrow \tilde{\nu}_1
h)/\Gamma(\tilde{\nu}_2\rightarrow \chi^0\nu)\sim
A^2 / ( m_{\tilde{\nu}_2}^2 {g_1}^2)$, and can be as large as $\sim
1/2$.  So in this case, the branching fraction for producing
$\tilde{\nu}_1 h$ could be $\sim 1/10$, leading to Higgs production rate 
of about ten times larger than that from gluon fusion (not just five times 
larger, since each strong 
production event gives two sparticles that can potentially produce a
Higgs).
In fact, in this case the rate of Higgs {\em pair} production through
$\tilde{\nu}_2$ decay is as large as the rate for $gg\rightarrow h$.
These pair production events would lead to striking final states
$b\overline{b}b\overline{b} llX$, with the invariant masses of both
$b\overline{b}$ pairs equal to $m_h$.  Note that even if
$m_{\tilde{q}}\sim m_{\tilde{g}}\sim 700$ GeV, the rate for cascade Higgs
production is only down from the $gg\rightarrow h$ rate by a factor of 
$\sim 10$, and would still likely allow for discovery because the
signal is not suppressed by the $h\rightarrow \gamma \gamma$
branching ratio.

Even more striking is the possibility that $\tilde{\nu}_2$ and
$\tilde{\nu}_1$ are the two lightest supersymmetric particles, with
$m_{\tilde{\nu}_2}>m_{\tilde{\nu}_1}+m_h$.  In this case, every squark
and gluino produced yields a Higgs particle in its cascade.
So for $m_{\tilde{q}} = 1.2 m_{\tilde{g}} =300$ GeV,
gluon fusion would account for only one in every $\sim$ 100 Higgs
particles
produced at the LHC.

Production of $\tilde{\nu}_2$'s and
their subsequent decay could also be an interesting source of Higgs
particles at the NLC.  The rate of Higgs production through $e^+ e^-
\rightarrow \tilde{\nu}_2 \tilde{\nu}_2$ is typically at least an order 
of magnitude lower than that due to $e^+ e^-
\rightarrow Z h $ and $WW$ fusion for $\sqrt{s}=500$ GeV
\cite{barnett, boos}.  However, $e^+
e^-
\rightarrow \tilde{\nu}_2 \tilde{\nu}_2$ could lead to sizable
Higgs {\em pair} production.  For example, for $\sqrt{s}=500$ 
GeV and $m_{\tilde{\nu}}=200$ GeV,
\begin{equation}
\sigma(e^+ e^- \rightarrow hh\tilde{\nu}_1 \tilde{\nu}_1)\simeq
6 \,{\rm fb} \cos^4\theta ({\rm BR}(\tilde{\nu}_2\rightarrow \tilde{\nu}_1
h))^2,
\end{equation}
compared to a cross section of $\sim .3-.5$ fb 
for the double Higgs-strahlung process $e^+
e^-\rightarrow Zhh$ in the decoupling regime, for $m_h\sim 100-130$ GeV
\cite{kilian}. Especially if
$\tilde{\nu}_2$ is lighter than all gauginos except for a bino-like state,
it is easy to choose $m_L$, $m_R$, and $A$ so that 
$\cos^4\theta ({\rm BR}(\tilde{\nu}_2\rightarrow \tilde{\nu}_1
h))^2$ is not more than an order of magnitude suppression (there is
even the possibility, as mentioned above, that that
 ${\rm BR}(\tilde{\nu}_2\rightarrow \tilde{\nu}_1 h)=1$).  Thus a
 possible signature of our scenario is an excess of events
with missing energy plus two $b \overline{b}$ pairs whose invariant
masses equal the Higgs mass, beyond the number expected from double
Higgs-strahlung followed by $Z\rightarrow \nu \overline{\nu}$.

\subsection{Invisible sneutrinos}

The motivation for considering this case in our scenario 
is that the $A$ terms suppress the masses of the lighter sneutrinos,
making it more likely than in standard schemes that some $\tilde{\nu}$'s
can either only decay invisibly, or not decay at all.  Let us assume
that $\chi_1^0$ and $\tilde{\nu}$ are the two lightest supersymmetric
particles. One immediate
question is whether the clean trilepton signal from
$\chi_1^{\pm}\chi_2^0$ production at
hadron machines remains, since $\chi_2^0$ has access to the invisible 2
body
decay mode $\tilde{\nu} \nu$. However, provided
$m_{\chi_2^0}>m_{\tilde l}$, $\chi_2^0$ can also decay through the
visible two body mode $\tilde{l} l$.  If the branching ratio for this
decay is not too small, the trilepton signal survives, because
$\chi_1^{\pm}$ decays into $\tilde{\nu}l$ and possibly $\tilde{l}\nu$, if
the
later is kinematically accessible.  Especially interesting is the
particular case where only the lightest $\tilde{\nu}$ mass eigenstate
is lighter than $\chi_1^{\pm}$, and
$m_{\chi^{\pm}_1}<m_{\tilde{l}}<m_{\chi_2^0}$.  The flavor of the
lepton produced in $\chi_1^{\pm}\rightarrow \tilde{\nu} l$ is
correlated with the flavor  of the light $\tilde{\nu}$. 
Moreover, it is reasonable to  expect the
lightest  sneutrino to be coupled to the largest $A$ term, and so,
assuming that the 
flavor structure for the $A$ terms resembles that of the neutrino
masses (the connection
between the two is most immediate in the sDirac case), the lightest
$\tilde{\nu}$ is likely to be mixture of $\tilde{\nu}_\mu$ and
$\tilde{\nu}_\tau$.  In this case, $\sim 1/2$ of the leptons produced
in the $\chi^{\pm}_1$ decays are $\mu$'s, while very few $e$'s are
produced, leading to roughly 7 $\mu$'s for every 4 $e$'s in the
trilepton signal, assuming that the $\chi_2^0$ decays produce equal
numbers of each lepton flavor.  

An invisible $\tilde{\nu}_1$ state can be produced along with a
heavier, visibly decaying $\tilde{\nu}_2$ at $e^+e^-$ colliders
through 
$s$-channel $Z$ exchange (and $t$-channel chargino exchange for
$\tilde{\nu}_e$), provided that $\sin 2\theta$ is not too small.
The decays of the heavier $\tilde{\nu}$
would lead to the signal $2l+{\not\!\!E}_T$ at the NLC. If
the masses of heavier, visibly decaying sneutrinos have already been
established through pair production, it should be possible 
to measure the mass of a light,
invisible $\tilde{\nu}$ using the energy
spectrum endpoints for the leptons produced in this process.  

As mentioned above, if $m_{\chi_2^0} < m_{\tilde l}$,
then the only two-body decay for $\chi_2^0$ is into $\tilde{\nu}\nu$,
so that
both $\tilde{\nu}$ and $\chi_2^0$ decay
invisibly.  The process
$e^+ e^- \rightarrow \gamma + {\not \!\! E}_T$ has been shown to be a 
feasible means for detecting the presence of these extra carriers of 
${\not \!\! E}_T$ at the NLC \cite{datta}.
 
Finally, suppose that $\tilde{\nu}$ and $\tilde{l}$, rather than
$\chi_1^0$ and  $\tilde{\nu}$, are the lightest supersymmetric particles.
Thus,
$m_{\tilde l}<m_{\chi_1^0}$, and $\chi^0_1$ decays visibly into
$\tilde{l}l$.  Meanwhile, the NLSP $\tilde{l}$ has only three-body decays, 
into $\tilde{\nu} l' \nu'$
and $\tilde{\nu}jj$.  In this case, a signal for slepton pair
production is $ljj{\not \!\!E}_T$, a characteristic signature for
chargino pair production (although the two cases are distinguishable
by their angular distrubutions, for instance) \cite{degouvea}.

\section{Flavor Changing Signals}
\label{sec:lfv}
In our framework, lepton flavor violating  contributions to $m_L^2$
can arise at tree level due to the same spurion(s) responsible for the 
Dirac neutrino masses and $A$ terms.  
This possibility exists in each of the three scenarios
discussed in section \ref{sec:bigsec}, but the connection between the
flavor structure of $m_L^2$ and that of the neutrino masses is most
direct for the sDirac case, which we therefore consider here for
simplicity.

Suppose that $X$, a chiral superfield
with $\langle X \rangle =\theta^2 F_X$, has the
appropriate flavor structure to induce Dirac neutrino masses via
\begin{equation}
{1 \over M^2}\d{ \, L X^{\dagger} N H_{u}}.
\end{equation}
Then one can also write down the Lagrangian term
\begin{equation}
{1 \over M^2} \d{(L^{\dagger}X)(X^{\dagger}L)},
\label{eq:lfvmsq}
\end{equation}
giving potentially large lepton-flavor violating 
contributions to $m_{L}^2$. As discussed in section \ref{sec:flavstruc},
$X$ could alternatively be a product of multiple spurions, some of
which conserve flavor and break supersymmetry, and others which do the 
opposite.  There are additional contributions to $m_{L}^2$ from the
spurion that generates the $A$ terms.  We will assume that the
flavor structure of this spurion is identical to that of $X$: this is
especially likely in the case that the supersymmetry-breaking piece of $X$ is
flavor conserving.

The contributions of (\ref{eq:lfvmsq}) might lead, for
example, to slepton oscillation signals at the NLC \cite{acfh} or, as considered
here, to rare lepton decays. 
Of course, similar contributions arise in more standard schemes as 
well:  a flavor breaking spurion $\frac{\langle \phi \rangle}{M} =\lambda$ that
generates Majorana neutrino masses through 
${1 \over M}[\lambda_{ij} L_i H L_j H]_F$ can
also appear in ${1\over M^2}[L^\dagger \lambda^\dagger 
\lambda L Z^\dagger Z]_D$, where $Z$ breaks supersymmetry but not
flavor.  
If the lepton-flavor violating contributions to $m_L^2$ are not
screened by much larger universal contributions, 
then a generic flavor structure for
$F_X$ in (\ref{eq:lfvmsq}), or for $\lambda$ in the standard case, 
leads to unacceptably large flavor violating signals.  On the other hand, 
not every structure for $F_X$ or $\lambda$ leads to a phenomenologically acceptable
neutrino mass matrix.  In light of this, we briefly
consider forms for $F_X$ motivated by neutrino
phenomenology, and estimate the flavor changing signals they induce.
We will not have specific flavor symmetries in mind that motivate the
textures we will consider.  Moreover we will estimate only the lepton
flavor violating signals due to the non-universal contributions to
$m_L^2$, and make the simplifying assumption that all other potential sources of
flavor violation (for instance, an $A_e$ matrix that is not aligned
with $\lambda_e$) vanish.

For our discussion we will assume that one Dirac neutrino is
hierarchically heavier than the others, and we will also take
there to be three $N$ states, although this is not an important 
assumption. Using our freedom to choose a basis for the $N$'s, we
consider the leading order flavor structure 
\begin{equation}
F_X \sim \pmatrix{0 & 0 & 0 \cr 0 & 0 & 0 \cr 0 & \alpha & \beta},
\label{eq:Fzero}
\end{equation}
with $\alpha$ and $\beta$ comparable.  That is, there is large
$\nu_{\mu}-\nu_{\tau}$ mixing as indicated by Super-Kamiokande, 
but $\nu_{e}$ has only a small component in the heavy state,  
to satisfy the CHOOZ bound \cite{CHOOZ}.

Since we assume that (\ref{eq:lfvmsq}) gives contributions
to $m_{L}^2$ as large as the universal ones, 
the form taken for $F_X$ suggests that 
the 23 entry of $m_{L}^2$ will be 
comparable in size to the diagonal entries.  
In this case
the branching ratio for the process $\tau \rightarrow \mu \gamma$
is near the current experimental limit for slepton masses near 100 GeV
\cite{gabbiani, feng}.  

Another potential signal is $\mu \rightarrow e\gamma$.  
The size of the branching ratio depends on the higher
order contributions to $F_X$ and is highly model dependent.  In the
abelian flavor symmetry models considered in \cite{feng}, with right
handed neutrinos integrated out above the flavor scale, both large
angle MSW and vacuum oscillation solutions to the solar neutrino
problem generally lead to too large a rate for $\mu\rightarrow
e\gamma$, and even models compatible with the small angle MSW solution
force the slepton masses above $\sim$ 500 GeV.  Here we do not
construct flavor models for light Dirac neutrinos but instead simply consider
the texture 
\begin{equation}
F_X \sim \pmatrix{\epsilon & \epsilon & \epsilon \cr \epsilon & \epsilon &
\epsilon \cr \epsilon & 1 & 1},
\label{eq:Fone}
\end{equation}
with only the order of magnitude of each entry, and not its precise
value, indicated.  One finds in this case that
$F_X^{\dagger} F_X$ has eigenvalues $\sim \epsilon^2$, $\epsilon^2$,
and $1$, and mixing angles $\theta_{23} \sim 1$,
$\theta_{13} \sim \epsilon$, and $\theta_{12} \sim 1$.   
Thus this case is
most likely to correspond to either large angle MSW or vacuum oscillation
solutions to the solar neutrino problem.  For vacuum oscillations,
we take $\epsilon^2\sim \Delta
m^2_\odot/\Delta m^2_{atm} \sim 10^{-7}$.  Since the 12 and 13 entries 
of $F_X^{\dagger} F_X$ are both of order $\epsilon$, we obtain
the order of magnitude relation
\begin{equation}
{B(\mu\rightarrow e\gamma)\over 1.2\times 10^{-11}} \sim .03 \left({100
{\rm GeV} \over \tilde{m} }\right)^4,
\label{eq:LFV1}
\end{equation}
where $\tilde{m}$ is a typical slepton mass, the lightest neutralino
is taken to be photino-like, and $m_{\tilde{\gamma}}^2/\tilde{m}=.3$
\cite{gabbiani, feng}.
This case is thus safe as far as $\mu\rightarrow e\gamma$ is concerned.
Depending on the assortment of assumed order unity factors we have
ignored, it is still possible, for light sparticle masses,
that the branching ratio will be accessible to future experiments .

On the other hand, for the large
angle MSW solution we should take $\epsilon^2  
\sim 10^{-2}$, leading to 
\begin{equation}
{B(\mu\rightarrow e\gamma)\over 1.2\times 10^{-11}} \sim \left({700
{\rm GeV} \over \tilde{m} }\right)^4,
\label{eq:LFV2}
\end{equation}
where we again take $m_{\tilde{\gamma}}^2/\tilde{m}=.3$.
Thus $\mu \rightarrow e\gamma$ forces the sparticle masses to be heavy.
One should keep in mind that these estimates have been obtained
ignoring other possible sources of lepton flavor violation, 
and for a particular texture for $F_X$.

One choice for the higher order entries in $F_X$ that is compatible
with the small angle MSW solution to the solar neutrino problem is
\begin{equation}
F_X \sim  \pmatrix{\epsilon^2 & \epsilon & \epsilon \cr
  \epsilon^2 
& \epsilon&\epsilon
\cr \epsilon^2 &1  & 1},
\end{equation}
leading to eigenvalues for $F_X^{\dagger} F_X$ $\sim \epsilon^4$,
$\epsilon^2$,
and $1$, and mixing angles $\theta_{23} \sim 1$,
$\theta_{13} \sim \epsilon^2$, and $\theta_{12} \sim \epsilon$.
Choosing $\epsilon \sim 1/30$ leads to acceptable masses and mixings.
Since the 12 and 13 entries of $F_X^{\dagger} F_X$ are order $\epsilon^2$, 
$B(\mu\rightarrow e\gamma)$ is suppressed by roughly a factor of $10^4$
relative to the large angle MSW case.

If large universal contributions to $m_L^2$ are present, they will
suppress the effects of
the flavor violating contributions induced by $F_X$ at tree
level.  However, even if we ignore (\ref{eq:lfvmsq}) entirely and
take a universal form for $m_L^2$ at, say, the GUT scale, we still obtain
potentially interesting flavor violating signals. This is because the $A$ terms
generate non universal contributions to $m_L^2$ radiatively:
\begin{equation}
\delta {m_L^2} ={1\over 8 \pi^2} A^\dagger A
\log(M_{GUT}/M_{SUSY}). 
\label{eq:nonu}
\end{equation}
This effect has been studied in \cite{hisano, ellis} in the context of see-saw
theories with a high scale for the right-handed neutrinos.  In these models, 
the Dirac neutrino Yukawa couplings are
sizable and generate additional non-universal contributions, leading to
\begin{equation}
\delta {m_L^2} ={1\over 8 \pi^2}(A^\dagger A +3 
\lambda^\dagger \lambda m_{3/2}^2)
\log(M_{GUT}/M_{N})
\end{equation}
for a universal scalar mass $m_{3/2}$ and a right handed neutrino
scale $M_N$.  Notice
that while in our framework only the $A$ term contributions are present and not
those from the Yukawa couplings, the logarithm is larger than in
the see-saw case because
the right handed neutrinos remain in the effective theory down to low
energies.  

To calculate rates for flavor changing processes due to
(\ref{eq:nonu}) for a given set of MSSM parameters,  
one needs to know the $A$ matrix.  This ambiguity 
is at the same level as in \cite{ellis}, where the Dirac neutrino Yukawa
couplings are unknown: the size of the largest coupling is fixed
by atmospheric neutrino data only once the scale of the right
handed neutrinos is specified (the $A$ terms are chosen
proportional to the Yukawas, with the scale set by the universal
gaugino mass).  Setting $M_N=10^{13}$ GeV, 
the authors of \cite{ellis} considered 
textures for $\lambda$
based on abelian symmetries and found promising signals for $\mu
\rightarrow e\gamma$ and $\tau \rightarrow \mu \gamma$ for significant 
portions of parameter space.  If we take $A=\lambda m_{3/2}$ in (\ref{eq:nonu}), then
our framework yields very similar signals to \cite{ellis} for a given choice
of SUSY parameters and a given form for $\lambda$.

\section{Conclusions}
The relationship $m_\nu \approx v^2/M$, where $v$ is of the order of
weak scale and $M$ an ultraviolet cutoff,
has been tremendously 
successful in describing the small mass of the neutrino. Whether
arising from Planck-scale suppressed operators in an effective theory, or
from a particular realization such as the see-saw mechanism, this has
been generally interpreted to signify the presence of lepton-number
violating physics at the scale $M$, well above the reach of
laboratory high-energy physics. This belief is predicated upon the idea
that the coupling of the neutrino to the lepton-number violating
sector of physics is order one, as might be expected in a GUT
see-saw, for instance.

The likelihood of a light Dirac neutrino has been discounted for
decades. Given the observed mass scales for neutrinos
from solar and atmospheric neutrino data, we would need a Yukawa coupling 
at $O(10^{-12})$ or smaller which appears difficult to understand
when the smallest
known Yukawa, $\lambda_e$, is $O(10^{-5})$. 
There is a possibility to explain the needed small Yukawa coupling as a
consequence of a new flavor symmetry $G_F$ broken only very slightly.
In previous efforts, 
the factorization of the symmetry
group into $G_F\otimes SUSY$ has been extended to the factorization of 
the symmetry breaking itself: vevs which break $G_F$ are supersymmetry 
preserving, while vevs breaking SUSY ($F \sim m_I^2 \sim v \mpl$) are
$G_F$ conserving.  If this  
is not the case, however, the Yukawa coupling can have an additional 
suppression factors in powers of $m_{I}/\mpl \sim 10^{-8}$.

The lightness of the Higgs doublets in supersymmetric theories ($\mu 
\sim v$) suggests
that such a factorization of symmetry breaking is inadequate. 
There should exist an additional symmetry group $G$ which is broken in the
supersymmetry breaking sector. Given that the Higgs is kept light by
$G$, we may ask whether there might be other
particles such as right-handed neutrinos, singlet under the standard
model, also kept light by $G$. 

\begin{table}[t]
 \renewcommand{\arraystretch}{1.2}
 \newcommand{\lw}[1]{\smash{\lower2.ex\hbox{#1}}}
 \begin{center}
  \begin{tabular}{|c||c|c|c|c||c|c|} \hline 
 &\multicolumn{4}{c}{See-Saw Theories}  &
\multicolumn{2}{c}{  Non  See-Saw Theories} \vline \cr \hline \hline
Mass scale & Conventional & & sMajorana & & Conventional & sDirac \cr
($M_{Pl}=1$)& see-saw     & &           & & EFT &  \cr
\hline
$1$ & $m_N$ & & & & & \cr
$m_I$ & & $m_N$ & & & & \cr
$m_I^2=v$ & $m_D$ & & $m_N$ && & \cr
$m_I^3$ & & &$m_D$ & $m_N$ & &\cr
$m_I^4 = v^2$ & $m_{LL}$ & $\times$ & $m_{LL}$ & $\times$ & $m_{LL}$ &
$m_D$ 
\cr 
\hline  \hline
\end{tabular}
\end{center}
\caption{Possible scenarios which achieve $m_\nu = v^2/M$. We only
  allow the various couplings to take values in powers of the
  intermediate scale $m_I\approx 10^{11}\gev$, as would occur if the
  couplings were generated in the supersymmetry breaking sector. $m_N$ is
  the Majorana mass for the right-handed neutrino, $m_D$ is the Dirac
  mass coupling the left- and right-handed neutrinos. $m_{LL}$ is the
  left-handed neutrino Majorana mass. $\times$ indicates that
  $m_{LL}$ cannot occur at $v^2$ for the given $m_N$. }
\label{tb:scenarios}
\end{table}

One immediate consequence is that the physics responsible for the
relationship $m_\nu\approx v^2/M$ is not occuring at the scale $M$,
and, in particular, that the mass of the standard model singlet state may be
much lighter than $M$ --- even as light as $m_\nu$ itself. The numerous 
new possibilities, employing only vevs in integer powers of
$\sqrt{F}=m_I$, are summarized in Table \ref{tb:scenarios}.  In
particular, there are interesting possibilities of sDirac (a light
Dirac neutrino) and sMajorana (a light Majorana neutrino with a weak-scale
right-handed neutrino) scenarios in a single generation.

The idea that $G$, which protects $m_N$, is broken in the supersymmetry
breaking sector is by no means purely philosophical. In theories in
which $G$ is broken by a supersymmetry preserving vev, we typically
expect {\em  all} couplings of $N$ supermultiplet to be highly suppressed.  
On the other hand, the case $G$ 
is broken in the supersymmetry breaking sector immediately invites the 
possibility of unsuppressed $A$ terms for right-handed scalar
neutrinos, which radically alter the  
phenomenology of this scenario relative to previous ones.

For instance, Higgs physics can be modified drastically, both in
production and decay. The mass spectrum of sleptons is dramatically altered 
and the presence of a light $m_{\tilde \nu}<45 \gev$ sneutrino is
permitted. Collider signatures can be changed dramatically. 
With or without lepton number violation, the sneutrino is
revived as a dark matter candidate. All of these things are easily
realized if $G$ is broken by supersymmetry
breaking vevs.

Furthermore, if the supersymmetry breaking sector breaks flavor symmetries
that are also broken in a separate supersymmetry conserving sector, we 
have a potentially new understanding of the large mixing in the
neutrino sector. Since the vevs of the fields $X$ which break
supersymmetry need not be aligned with flavor violating vevs that
preserve supersymmetry, there is 
no reason to expect the mass eigenstates of neutrinos to be aligned
with those of charged fermions, although they may still have a
hierarchical structure.

When viewed from the perspective of the $\mu$ problem, such a scenario 
is exceedingly natural. The presence of unsuppressed $A$ terms provide not
only exciting phenomenology, but also the promise that this scenario
can be tested in the near future. While experiments will provide the
ultimate test of these ideas, this framework provides exciting
possibilities for connections between what previously have seemed
separate elements of supersymmetric theories.

\vskip 0.25in
\appendix
\noindent {\bf \Large Acknowledgements}
\vskip 0.15in
\noindent This work was supported in part by the Director, Office of Science,
Office of High Energy and Nuclear Physics, Division of High Energy
Physics of the U.S. Department of Energy under Contract
DE-AC03-76SF00098 and in part by the National Science Foundation under
grant PHY-95-14797.

\end{document}